# Switching of the Mott transition based on the hole-driven MIT theory

Hyun-Tak Kim (htkim@etri.re.kr), Bong-Jun Kim, Yong Wook Lee, Byung-Gyu Chae, Sun Jin Yun

Convergence & Components Lab., MIT Device Team, ETRI, Daejeon 305-350, Korea

Switching voltage of first-order metal-insulator transition (MIT) in VO$_2$, an inhomogeneous strongly correlated system, is changed by irradiating an infrared light with wavelength, 1.5 micrometer, and applying the electric field (photo-induced switching). This was predicted in the hole-driven MIT theory in which hole doping of a low concentration below 0.01% into conduction band (Fermi surface) induces the abrupt MIT as correlation effect. The switching is explained by the Mott transition not the Peierls transition.

The argument whether a well-known representative half-filling insulator VO$_2$ (paramagnetic, V$^{4+}$, 3d$^1$) is the Mott insulator or the Peierls insulator has been made for a long time [1,2]. This was due to observation of the structural phase transition (SPT) from monoclinic to tetragonal near 68$^o$C regardless of correlation effect, although the correlation effect is the first-order metal-insulator transition (MIT) jump caused by breakdown of a critical on-site Coulomb interaction, U$_c$ [3-5].

In order to reveal the mechanism of the first-order MIT, we have developed the hole-driven MIT theory (extended Brinkman-Rice (BR)) for an inhomogeneous system [4-6] and experimentally observed the first-order MIT not undergoing the SPT and a MCM (monoclinic and correlated metal) phase in VO$_2$ [7]. These are evidence of the Mott first-order MIT.

This paper shows a new observation of a switching phenomenon of the first-order MIT predicted in the hole-driven MIT theory [3,4] and clearly explains the correlation effect. The switching behavior is a strong evidence of the Mott transition not the Peierls transition accompanied with a structural distortion.

The MIT switching was performed by irradiating an infrared light with wavelength ¥ë 1.5 μm and applying electric field(voltage) on a VO$_2$-based two-terminal device. The beam diameter is about 5 μm. The fabrication method of the device was explained in a previous paper [6]. The thickness of VO$_2$ films was approximately 100 nm. The films were deposited on Al$_2$O$_3$ substrates. The dimension of device I in Fig. 1(c) was the distance, L=5 μm, between two electrodes and the width, W=50 μm, as shown in the inset of Fig. 1(c). The dimension of device II for a MIT switching in Fig. 1(d) was L=30 μm and W=3 μm.

For an inhomogeneous system with a local metal phase and a local Mott insulator phase in a measurement region, the measured conductivity has an averaged value in which carriers in the local metal phase are averaged over lattices in the measurement region. The averaged system has an electronic structure of one effective charge per atom [4,5]. This is the same electronic structure as one of the BR picture [3]. The averaged effective mass of quasiparticle [4,5] has the same form as that in the BR picture and is given by

m$^*$/m=1/(1-(U/U$_c$)$^2$),

$\qquad$ =1/(1-κ$^2$ρ$^4$), $\qquad$ for U/U$_c$=κρ$^2$ $\qquad$ in [4,5]

$\qquad$ =1/(1-ρ$^4$), $\qquad$ at κ=1 in [4,5]. $\qquad$ (1)

The band filling, ρ, is the content of the metal phase in the inhomogeneous system. Eq. (1) is defined at ρ≠1 (Fig. 1(a)) and has a first-order discontinuous MIT between a Mott insulator with U$_c$ at ρ=1 and a metal at ρ$_{max}$<1 (yellow part in Fig. 1(a)). The MIT is caused due to breakdown of U$_c$ by hole doping of a very low density, n$_c$=1-ρ$_{max}$, into the Mott insulator (Fig. 1(a)); this is a hole-driven MIT. n$_c$ is a minimum hole density where the MIT occurs. Conversely, control of n$_c$ makes the Mott insulator switch between insulator and metal. After the MIT, the local Mott insulators become strongly correlated local metals with a correlation strength, κ≠1, in the BR picture. m* in Eq. (1) is an average of the true effective mass in the BR picture [3]. Further, the correlated metal is regarded as a non-equilibrium state because metal exists at the divergence in Eq. (1) (Fig. 1(a)).

Fig. 1(b) shows the temperature dependence of resistance with a resistance jump near 60$^o$C for VO$_2$; this is well known result. The inset is the temperature dependence of carriers measured by Hall effect [6]. The type of carrier changes from hole in semiconductor to electron in metal near 59$^o$C. The number of the measured hole carriers is 1.16 x 10$^{17}$ cm$^-$3 at 59$^o$C. We can guess the number of hole carriers is less than at most n ≈ 1.0 x 10$^{18}$ cm$^{-3}$ which corresponds to 0.006% of the number of carriers in the half-filled band, when one electron in the cell volume, 59.22 x 10$^-$24 cm$^3$, of VO$_2$ is assumed. Although we assume the maximum number of hole carriers, n$_c$ ≤ 0.01% can be deduced. The Hall data indicates that hole doping of a low concentration into conduction band induces electron carriers of a high concentration through MIT. This can be explained by breakdown of U$_c$ [5].

Fig. 1(c) shows the current vs voltage characteristic for device I. Current increases exponentially just before the abrupt jump, as shown in the inset, and linearly after the jump with increasing voltage. The exponential behavior is the semiconducting characteristic with hole carriers. The linear Ohmic behavior is the metallic characteristic and an extrinsic phenomenon due to resistance (electron-phonon interaction in metal), and results in Joule heating followed by the SPT. The current density in the Ohmic region is higher than $1 \times 10^6$ A/cm². This indicates that VO₂ became metal, and that electric field excited the hole charges of $n_c$ into conduction (or valence) band (Fermi surface). The jump is the intrinsic correlation characteristic. The magnitude of the jump is maximum (divergence in Eq. (1); maximum relaxation time) at T=0 K where phonon does not exist (no Ohmic behavior). As the measurement temperature increases, the magnitude of jump decreases due to the decrease of resistance [6]. This is because the jump is screened (see Fig. 12 of Ref. 6). When the electron-phonon interaction (resistance) in metal exists at an arbitrary temperature, it is very difficult to observe the correlation effect because the effect of electron-phonon interaction is larger than that of electron-electron interaction. Note that the jump as correlation effect does not exist in the Mott-Hubbard (MH) model which shows the continuous MIT with a change of the Coulomb interaction. The MH model well describes the antiferromagnetic insulator.

Fig. 1(d) shows current-voltage curves observed by incidence of an infrared light of ¥e1.5 µm into the device II. As the light intensity increases, the MIT voltage decreases. The current (at red arrow shown in Figure) in which the MIT occurs is the same at all transition points. This indicates that the number of hole carriers is the same at the transitions because light make hole charges to induce hole carriers, and that the VO₂ film at the transition points does not undergo the SPT [6,8]. Thus, a photo(light)-induced MIT switching occurs at $n_c$ and the light plays a gate role in a switching transistor.

The MIT (jump) is caused by $n_c$(Light,E) = n(E) + n(Light), where n(E) is the hole density excited by electric field (voltage), n(Light) is the hole density excited by light, $n_c$(Light,E) is the critical hole density in which the MIT occurs [7]. For constant $n_c$, n(Light) decreases as n(E) increases. Thus, the MIT voltage changes, which is evidence of the fact that the MIT is simultaneously controlled by holes induced by electric field and light. This was predicted in Eq. (1).

In conclusion, the first-order MIT occurs by hole injection. Since the SPT occurs only at $T_{SPT}$, the variable MIT voltage (switching) is explained in terms of not the Peierls but the Mott picture; this indicates VO₂ is the Mott insulator. (Shown on May 13, 2007, SCES'07)

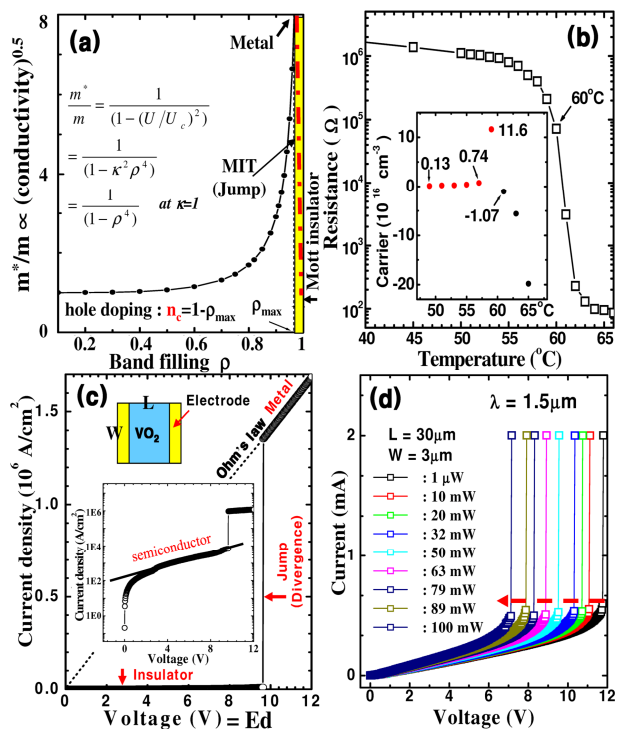

Fig. 1(a) Hole-driven MIT theory. Yellow part is the first-order jump. m*/m ∝ σ⁰·⁵, σ is conductivity. (b) The temperature dependence of resistance measured with a VO₂ film. Inset shows the temperature dependence of carriers measured by Hall effect. Hole carriers are red solid circle and electron carriers are black solid circle. (c) Electric field-induced MIT measured with device I. Inset is log graph. (d) Infrared power dependence of the MIT voltage measured at room temperature with device II. The movement of the MIT voltages is due to switching. Infrared light plays a gate role in the MIT switching.

## References

[1] T. M. Rice, H. Launois, J. P. Pouget, Phys. Rev. Lett. 73 (1994) 3042.

[2] R. M. Wentzcovitch, W. W. Schulz, and P. B. Allen, Phys. Rev. Lett. 72 (1994) 3389.

[3] W. F. Brinkman and T. M. Rice, Phys. Rev. B2 (1970) 4302.

[4] H. T. Kim, Physica C 341-348 (2000) 259; 'New Trends in Superconductivity' NATO Science Series II Vol. 67 p137 (Eds, J. F. Annett and S. Kruchinin, Kluwer, 2002), cond-mat/0110112.

[5] H. T. Kim, B. J. Kim, Y. W. Lee, S. J. Yun, and K. Y. Kang, Physica C 460-462 (2007) 1076.

[6] H. T. Kim, B. G. Chae, D. H. Youn, S. L. Maeng, G. Kim, K. Y. Kang, and Y. S. Lim, New J. Phys. 6 (2004) 52.

[7] H. T. Kim, Y. W. Lee, B. J. Kim, B. G. Chae, S. J. Yun, K. Y. Kang, K. J. Han, K. Yee, and Y. S. Lim, Phys. Rev. Lett. 97 (2006) 266401.

[8] H. T. Kim, Y. W. Lee, B. J. Kim, B. G. Chae, S. J. Yun, and K. Y. Kang, cond-mat/0603546.